\documentclass[11pt]{article}

\usepackage{graphicx}

\title{Open Systems' Density Matrix Properties in a Time Coarsened Formalism}

\author{Robert Englman$^{a,b}$ and Asher Yahalom$^{b}$  \\
$^a$Soreq NRC, Yavne 81800, Israel\\
$^b$Ariel University, Ariel 40700, Israel\\
e-mail: englman@vms.huji.ac.il; asya@ariel.ac.il; }

\begin{document}

\newcommand{\beq} {\begin{equation}}
\newcommand{\enq} {\end{equation}}
\newcommand{\ber} {\begin {eqnarray}}
\newcommand{\enr} {\end {eqnarray}}
\newcommand{\eq} {equation}
\newcommand{\eqs} {equations }
\newcommand{\mn}  {{\mu \nu}}
\newcommand{\sn}  {{\sigma \nu}}
\newcommand{\rhm}  {{\rho \mu}}
\newcommand{\sr}  {{\sigma \rho}}
\newcommand{\bh}  {{\bar h}}
\newcommand {\er}[1] {equation (\ref{#1}) }
\newcommand {\ern}[1] {equation (\ref{#1})}
\newcommand{\mbf} {{ }}
\newcommand {\Del} {\Delta}
\newcommand {\SE} {Schr\"{o}dinger equation}

 \maketitle
 \begin {abstract}
The concept of time-coarsened density matrix for open systems has
frequently featured in equilibrium and non-equilibrium statistical
mechanics, without being probed as to the detailed consequences of
the time averaging procedure. In this work we introduce and prove
the need for a selective and non-uniform time-sampling, whose form
depends on the properties (whether thermalized or not) of the
bath. It is also applicable when an open microscopic sub-system is
coupled to another {\it finite} system. By use of a time-periodic
minimal coupling model between these two systems, we present
detailed quantitative consequences of time coarsening, which
include initial state independence of equilibration, deviations
from long term averages, their environment size dependence and the
approach to classicality, as measured by a Leggett-Garg type
inequality. An interacting multiple qubit model affords comparison
between the time integrating procedure and the more conventional
environment tracing method.

\end{abstract}

PACS number(s): 03.65.Yz,75.10.Jm, 05.30.-d

\section{Introduction}
 Though not usually presented as such, the density operator
 formalism provides at least a partial resolution of the quantum mechanical
 time-arrow conundrum (time inversion invariant equations {\it
 versus} the expected and observed uni-directional trend towards equilibration), and does so by
 invoking some averaging procedure that spoils the unitary
 development of the pure state density matrix (DM). The equilibration issue has been particularly
 acute for {\it closed} systems (the Universe or part of it hermetically
 segregated from the rest, such as trapped ultracold atoms) and has featured in many articles both much
 cited ones \cite{Deutsch}-\cite{RigolDO} and those quite recent
 \cite {PopescuSW}-\cite{EblingKF}. The
 commonality among these appears to be the inclusion of a random
 process that is not implicit in the defining equation. There is
 no consensus on the dependence (or on its absence) on initial
 conditions of the equilibrating system and in what situation this (or that) occurs
 (e.g.,\cite{SteinigewegKNGG}).

 The time-arrow conundrum lends impetus to the present work which derives density matrices in
 arbitrary representation for {\it open} systems, such that a microscopic
 system is coupled to an ancilla (which can be another microscopic system or a finite or infinite
 "bath"), and does so by a method of time averaging of the microscopic system's state
 function, in contrast to other, more frequently employed
 procedures, in which  a stochastic process or tracing over the ancilla's states is
 invoked \cite {BreuerP}. It is a development arising from a previous work \cite
 {EnglmanY2013}, in which justifications of the time
 averaging method were given based on the ergodic hypothesis
 (equating time- and state sampling- averages) \cite {Farquhar, FalascoSV}, and the "minimal
 coupling model" (shown in \er{H0} below) was heuristically
 derived. [We note that this model, also known as the "monochromatic coupling model",  has quite a history starting
 with laser coupling \cite {AutlerT} and other topics, supported by  detailed theoretical analysis of the model in the
 context of Floquet's theorem by
 \cite {Shirley} and \cite{ShevchenkoAN}, and very recently in relation to
 Landau-Majorana-St\"uckelberg-Zener interferometry \cite {GaneshanBD}, also \cite{NavonKAGAO})

 After describing (for completeness and in a  nutshell) our simplified employment of the time averaged DM method (in section 2),
 the present work makes the
 following advances requisite for the establishment and practice of
 the method: It shows first (in section 3) that the method satisfies the basic theoretical conditions on
 DM. It then  identifies quantitatively the forgetting of initial conditions (section 4). In the short, but central section 5 (propped up by
 the formal proof in Appendix D) we delineate, apparently for the first time, how the "time coarsening" method for the DM
 is to be used for open systems coupled to finite-sized environments and, by a natural extension to infinite-sized
 environments, giving rise in general to {\it non-uniform time weighting}. Continuing, the paper develops an interacting N-qubit
 model (in section 6.1) in order to  show  how the standard
 deviation ("distance") of the open system DM
 vanishes with the size  of the ancilla. This leads us (in section 6.2) to a quantitative comparison between the
 ancilla tracing and time averaging procedures regarding the  ancilla's size dependence of the DM standard deviation in each procedure.
 The formalism is then applied in section 6
 to a Leggett-Garg type inequality to investigate quantitatively how the inequality,
  violated by quantum systems, is reinstated as the ancilla size approaches macroscopics.
 To make the paper self contained we have briefly reiterated in Appendixes A and B  the bare essentials of \cite{EnglmanY2013}.
As already noted, the way time averaging is to be done for an open
(microscopic) system  coupled to a  {\it finite} size environment
is the subject of the (mathematical) Appendix D, with implications
also for general macroscopic environments.

\section  {Time Averaging in the Minimal Coupling\- Model (MCM)}
This model was introduced in \cite {EnglmanY2013}, with
motivations and heuristic justification given there, in the form
of the Hamiltonian \beq H(t)= E\sigma_z+k\sigma_x \sin(\omega t)
~~~(\omega \to 1)\label{H0}\enq and the (transposed) solution of
the relevant time-dependent \SE~ with pre-fixed initial conditions
\beq\psi(t)^T= [\psi_u(t),\psi_l(t)]\label {psiul}\enq
$\sigma_{z,x}$ are Pauli matrices, $2E$ is the Zeeman splitting,
$k$ the coupling strength of the spin to the reservoir and the
time ($t$) dependent circular function represents the minimal
effect of the latter on the former; $u,l$ label the upper and
lower component in the spinor solution. The DM $\rho_{ij}$ in an
$ij$ representation is derived from the time averages over a time
window  $2\Delta t$ of  \beq M_{ij}(t)\equiv <i|\psi(t)><\psi
(t)|j>\label{M}\enq in the form \beq \rho_{ij} = \frac{1}{2\Delta
t} \int_{t-\Delta t}^{t+\Delta t}d\tau M_{ij}(\tau)~~ (i=u,l)
\label{rho1}\enq

\section {Necessary Conditions for Density Matrices $\rho_{ij}$}
These were formulated by Fano, as follows \cite{Fano}:
\begin{enumerate}
\item Matrix is Hermitian. This is clearly the case in the above
MCM. \item Matrix trace is one. For the time averages this follows
from the normalization to unity of the wave function at any time.
\item In any representation the diagonal elements $\rho_{ii} \geq
0$: this follows from the  definition.
 \item When diagonalized by a unitary transformation $Tr
\rho^2\leq 1$. This requires that for each diagonalized eigenvalue
$(DM)_n(\rho)$ of the DM,  $0\leq (DM)_n(\rho)\leq 1$, in which
the equalities hold for pure states. In the present spin-half
formalism with two DM eigenvalues this translates to \beq
0\leq\frac{1}{2}(1\pm \sqrt{G(t)})\leq 1 \label{Ineq}\enq where
the unity in the parentheses comes from unit trace. Here, with a
little manipulation, \ber G(t)&\equiv& [\frac{1}{2\Delta
t}\int_{t-\Delta t}^{t+\Delta
t}(|\psi_u(t')|^2-|\psi_l(t')|^2)dt']^2
\nonumber \\
 &+& |\frac{1}{\Delta t}\int_{t-\Delta t}^{t + \Delta t}\psi_u^*(t')\psi_l(t'))dt'|^2
\label{G} \enr (involving the upper ($u$) and lower ($l$)
components of the wave function) and this has to be numerically
less than $1$ for the last Fano requirement to hold. Now, $G(t)$
is clearly non-negative and is maximal when the two components are
throughout real and positive, and limits for this quantity need to
be sought. The derivation of a maximum of unity is given in
Appendix C.
\end{enumerate}

This completes the proof for the satisfaction of the Fano
requirements for time averaged DM.
\section{Equilibration}
The first question that needs to be asked about the model is how
it describes equilibration characteristics of the spin system.
(The mode and time-duration of the evolution {\it to} the
equilibrated state, discussed recently in e.g. \cite{GoldsteinHT},
is outside the scope of the present time-integrated approach.
Historically, the dependence of the rate of equilibration on the
power-spectrum of the coupling seems to have been first given in
\cite{KuboT}.) Defining the equilibrated value of $\rho_{ij}(t)$
as that value which is only minimally dependent on time (an issue
discussed further in the "Distance" section of this work), we ask
how does this depend on the initial conditions and on the strength
of the coupling $k$? Also, at what values  of this coupling and of
other parameters is "full equilibration" ($=$ the maximal entropy
stage) attained? Figures 2 and 3 show the results.
\begin{figure} \vspace{6cm} \includegraphics{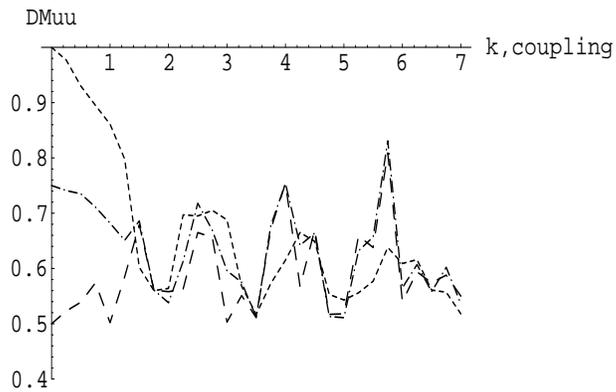} \caption {Upper diagonal density
matrix (DM)  as function of the coupling strength $k$ for the
half-Zeeman splitting $E=1$, and the following starting values:$~
1,~ 0.75, ~0.5.$ } \label{DME1}
 \end{figure}
 \begin{figure} \vspace{6cm} \includegraphics{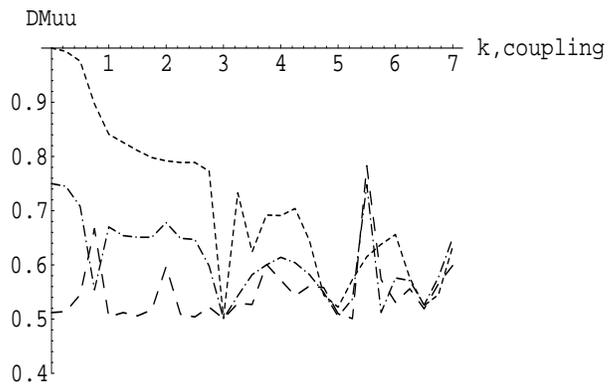} \caption {Upper diagonal DM  as
function of the coupling strength $k$ for a different half-Zeeman
splitting $E=\sqrt{2}$, and the following starting values:$~1,~
0.75,~ 0.5.$} \label{DMESqrt2}
 \end{figure}
 The curves show that the starting value  is forgotten beyond
 $k=2$ in figure 1  with $E=1$ and beyond about $k=3$
  in figure 2 with $E=\sqrt{2}$.

  Another important lesson from the figure is the strong fluctuations in
  the results, so that for physical interpretation only the
  rough averages and tendencies are meaningful. This strongly fluctuating property of
  the MCM and its analogues are well known from previous works
  \cite {AutlerT}-\cite {GaneshanBD},\cite{IrishGMS}.

  \section  {Non-Uniform Time-Weighting}
  \subsection {Finite Bath: Discrete Energy Levels}

 For a finite, though large, environment (bath) its level
 scheme is strictly discrete and the density of states is
 singular. The  coupled subsystem's instantaneous energy $E(t)$ is continuous.
 The two systems form together a pure state with eigenenergy
 $E_{total}$ (\cite {LindenPSW}).
 Then  the continuous (Lebesgue) time integration  in \er{rho1} with uniform time weighting has to be replaced by a discrete sum.
 Alternatively, the integral has to
  weighted by the singular factor  \beq
\sum_{i}\delta(E_{total}-E(t) -
E_i)/<\psi(t)|\delta(E_{total}-H_{system} -E_i)|\psi(t)>
\label{F3}\enq where the sum runs over all energy values of the
environment and $\delta(x)$ is the Dirac delta function. A proof
of this formula is given in Appendix D and an application is
provided in section 6.
\subsection {Macroscopic Bath: (Quasi-) Continuous Energy Density}
For this case, in a straightforward extension of the previous
expression, the  weighting factor for time coarsening is \beq
D(E_{total}-E(t))/<\psi(t)|D(E_{total}-H_{system})|\psi(t)>\label{F4}\enq
in which $D(E_{bath})$ is the distribution of bath energies. This
form was used in \cite {EnglmanY2013}, there based on heuristic
reasoning, for a bath distributed as in a canonical ensemble.
\section {"Distance" from equilibrium state}
Pursuing the environment-tracing formalism to derive the reduced
density matrix of a small subsystem, and starting with an
arbitrary but representative pure state of the subsystem {\it
plus} the environment, in their above quoted paper the Bristol
group have shown that the long term time averaged deviation
("Distance") of the subsystem state (its truncated density matrix)
from the equilibrated state is (or is upper-limited) proportional
to the square root of the ratio between the subsystem's dimension
in its Hilbert space and that of environment (Equation 8 in \cite
{LindenPSW}, also \cite{PopescuSW}). Before establishing an
analogous result for the time integration formalism we show that
an interacting $N$ half-spin model verifies numerically the
predicted inverse square root relationship for the "distance"
\cite{remark}.
\subsection{$N$-spin model}
Such systems with finite Hilbert state dimensions have been
extensively studied for entanglement and other properties. A
recently built processor consisting of eight interacting qubits
and subject to time dependent interaction was investigated as a
means to achieve feasible quantum computing \cite {LantingPS}.

 In our model the total system of $N$ $1/2$-spins (qubits), consisting
of one subsystem (labelled $0$) and ($N-1$) bath $1/2$-spins
($i=1...N-1$), interact according to a Hamiltonian \beq H(N)=
\sum_{i=0,...,N-1} E_i\sigma_{zi} + \sum_{i,j=0,...,N-1;i\neq
j}g_{ij}\sigma_{xi}\sigma_{xj}\label{HN}\enq with energy
parameters, so chosen as to avoid any degeneracies in the
eigenenergies, and two sets of coupling parameters for
$g_{ij}=g_{ji}$, given by  \beq E_0=1; E_j=1.5\sin^2\frac{2\pi
j}{10};~g_{0j}=2.5,3.5;~g_{ij}=.75,1.5;~~1\leq i,j\leq N-1\label
{param}\enq The matrix of this Hamiltonian, that has to be
diagonalized to obtain its eigenenergies and pure-state
eigenfunctions, is of a square dimension of $2^N$ x $2^N$. To
formally write out in the spin's representation the matrix  for
large $N$ and then let it be solved by a routine (e.g., Fortran or
Mathematica), we have used the following trick, not used (to our
knowledge) heretofore: We have replaced the ordinal, decimal
number of the matrix rows (or columns) in the routine by its
binary representation and have formalized the matrix elements in
this representation. Thus, there is an entry at the decimal
($13,25$), or ($25,13$) matrix position with the value of
$g_{3,5}=g_{5,3}$, because in the binary representation we have
for $N=8$ (with the spins labelled as $0,1,...,7$, in which  the
last $7$ digits label the bath spins) \beq
13=[000\bar{0}1\bar{1}01],~~~25=[000\bar{1}1\bar{0}01]\label{binary}\enq
with the $3$ and $5$ spins that flip and counter-flip identified
by a superbar.

\begin{figure} \vspace{6cm} \includegraphics{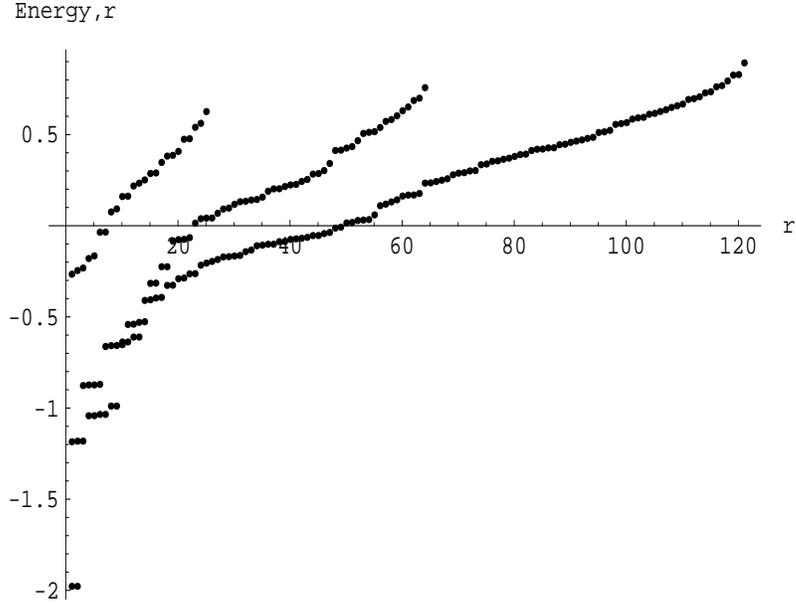} \caption {Eigenenergies (in
arbitrary units) against the serial number $r$ of increasing
energies in an interacting $N$-spin system for $N=5,6,7$}
\label{DMSpinEner}
 \end{figure}

The $2^N$ ordered energy values (labelled with the ordinal number
index $r$) obtained by numerical diagonalization of the
Hamiltonian \er{HN} are shown in Figure \ref{DMSpinEner}. The
figure shows sparseness in the lower range, uniform linear
increase in the middle and some super linear increase towards the
upper end.

The reduced density matrix of the small subsystem $\rho^S$ (of
size $2$ x $2$) is obtained in the bath-tracing procedure by
tracing over $2^{N-1}$ bath states the total system's density
matrix for any of its pure eigenstates (whose number is $2^N$).
The "distance" is taken as the root mean square deviation of
$\rho^S$ over the different choices from the $2^{N-1}$ bath
eigenstates.

\begin{figure} \vspace{6cm} \includegraphics{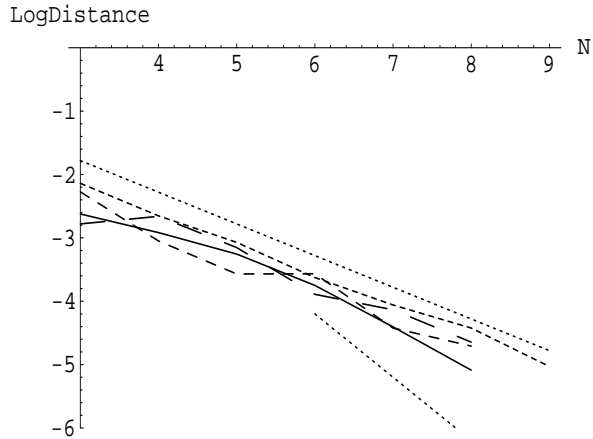} \caption {"Distances" {\it versus} N,
explained in the text. } \label{DMDist}
 \end{figure}

\subsubsection {N-spin distances}
In Figure \ref{DMDist} we show
our computed "Distances", plotted logarithmically (Log$_{10}$)
against the dimension of the Hilbert space of the ( $N -1$)
half-spins with which the system (also: a half-spin) interacts.
The full curve, the long broken line curve, the medium broken line
curve and the
 short broken line curves  are for different parameter sets
in the  Hamiltonian;  the  one (reaching up to $N=9$) is for a
parameter set  in  which all the $\sigma_x\sigma_x$ coupling
strengths  have the same value $2.5$ for every (bath-spin) -
(system-spin) coupling $g_{0i}$ and the value of $0.7$ for all
inter- bath couplings $g_{ij}$. The other curves are for varied
coupling strength of the  same order. The energy scale is set by
the Zeeman-splitting ($=2$) for the  subsystem; the bath spins
have  varying splittings of similar magnitudes.

The two straight, dotted lines bordering the computed curves show
 putative dependencies of the "Distance" on the Hilbert space
dimension of $2^{(N-1)}$, the
 lower dotted line following (decimal logarithmically) the inverse of this dimension
 and the upper dotted line the inverse square root. Asymptotically, the
computed values appear to follow the inverse square root law, in
line with the predicted upper limit dependence in Eq. (8) of \cite
{LindenPSW}.
\subsection{DM through time summation with random partitioning}
\begin{figure}
\vspace{8cm}
\includegraphics{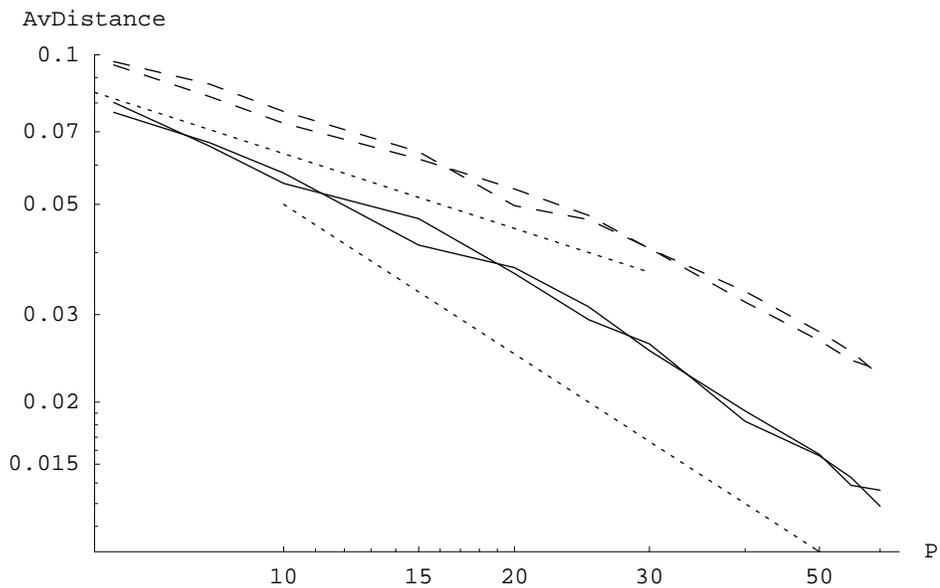} \caption {Distance (or Standard Deviation SD of the
density matrix upper eigenvalue over a large time interval)
against the number P of randomly partitioning  the time
integration interval $ 2\Delta t=4\pi$ and summing the value of
the upper eigenvalue of the density matrix at the midpoint of each
partition. The curves are for parameter values of the Hamiltonian
shown in \er{H0}
 $\omega =1,e=0.5$. The solid lines are for $k\approx 1$ and
the broken lines $k=1.5$, with two different random partitioning
in each case. The two dotted straight lines are, in the
logarithmic plotting of the figure, slopes of $P^{-\frac{1}{2}}$
and $P^{-1}$, respectively.
 }\label{DMSDRandom1}
 \end{figure}

As noted in section 5 and Appendix D, for a finite-sized ancilla
the time integration over a window of $2\Delta t$ needs to be
replaced by a discrete time-summation. We have subdivided the
window into P segments (partitions) having {\t randomly} chosen
relative lengths and summed the DM eigenvalues at P points
corresponding to the mid points of the segments. Figure \ref
{DMSDRandom1} shows a characteristic set of results \cite
{remark2}. For finite P the discrete summation corresponds to a
finite bath while for $P\to\infty$ the Standard Deviation (SD)
should approach the time {\it integration} value, corresponding to
an infinite bath environment. The computed curves appear to show
initially a $P^{-\frac{1}{2}}$ relationship, similar to the
inverse square root dependence for the environment tracing result
in figure \ref{DMDist},  followed for large $P$ by one like
$P^{-1}$.

The two figures, \ref {DMDist} and \ref{DMSDRandom1}, show the
congruity of the environment tracing and the time integration
methods for a general result in \cite{LindenPSW} which is ${\it
model~ independent}$.

 ~

 ~

\section{A Leggett-Garg-Wrachtrup (LGW) Inequality}
The Bell inequalities for correlations between measurements at
different locations, famously potentially violated by quantum
systems while observed by classical systems, were formulated for
temporally subsequent measurements by Leggett and Garg \cite
{LeggettG}. A review of experimental and theoretical developments
is found in \cite{EmaryLN}. A modification which tests the system
for  "reality" ( meaning the characterization of the physical
system, irrespective of whether it is measured or not) and
"stability" ("the conditional probability $Pr(i,t|i,t_0)$
 to find the system at times $t,t_0$ in the same state $i$,
  provided that this probability depends only on the time
difference") was proposed in \cite{WaldherrNHJW} by an inequality
written, with $t_0=0$ being the starting time and $t$ a subsequent
instance, as \beq Pr(i,2t|i,0)-Pr^2(i,t|i,0)\geq 0\label{LGW}\enq
 We use this inequality to answer two questions:
\begin{enumerate}

  \item Does the (reduced) density matrix arising from an open microscopic quantum system interacting with
 an ancilla have the property of a quantal or a classical system,
 in the sense of violating or observing the above inequality? (We
 recall that in a measurement of a quantal observable, after
 interaction with the measuring apparatus, the truncated DM
  has the status of classical probabilities.)

 \item Assuming a classical status for the DM of an open system
 linked to an "infinite" environment  and a quantum status for the
 same system when linked (entangled) to another microscopic
 system, is there a way to decide between the
status of the combined systems in terms of the requirements of the
above LGW inequality?
\end{enumerate}
It is proposed that the time integrating formalism leads to
answers in the following sense: The execution of the time
integration is dependent on the coupled ancilla. For a macroscopic
ancilla this has to be carried out over an effectively full
temperature range (or over a full period, if there is a
periodicity in the system), whereas for a finite ancilla the
integration is to be restricted, either to a summation or to a
limited range of the integration. We then test the violation of
the LGW inequality by increasing the time averaging range $2\Delta
t $ from zero (the pure state case) to its effectively full
averaging value $2\Delta t=4\pi$, appropriate to a macroscopic
bath.

\begin{figure} \vspace{8cm} \includegraphics{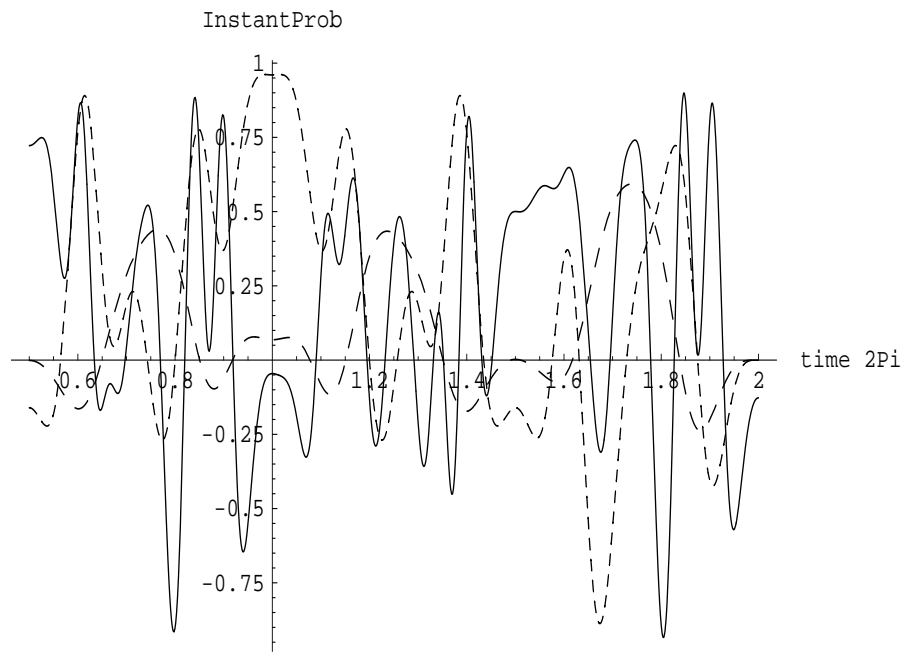} \caption {Pure sate results for the
LHS of \er{LGW2} obtained from solutions of the time dependent
\SE~ with a choice of parameters, each choice drawn with different
lines. The curves are mainly positive, but there are some times at
which they are negative, thereby violating the inequality in
\er{LGW2}} \label{DMLGAIInstProb}
 \end{figure}

\begin{figure} \vspace{6cm} \includegraphics{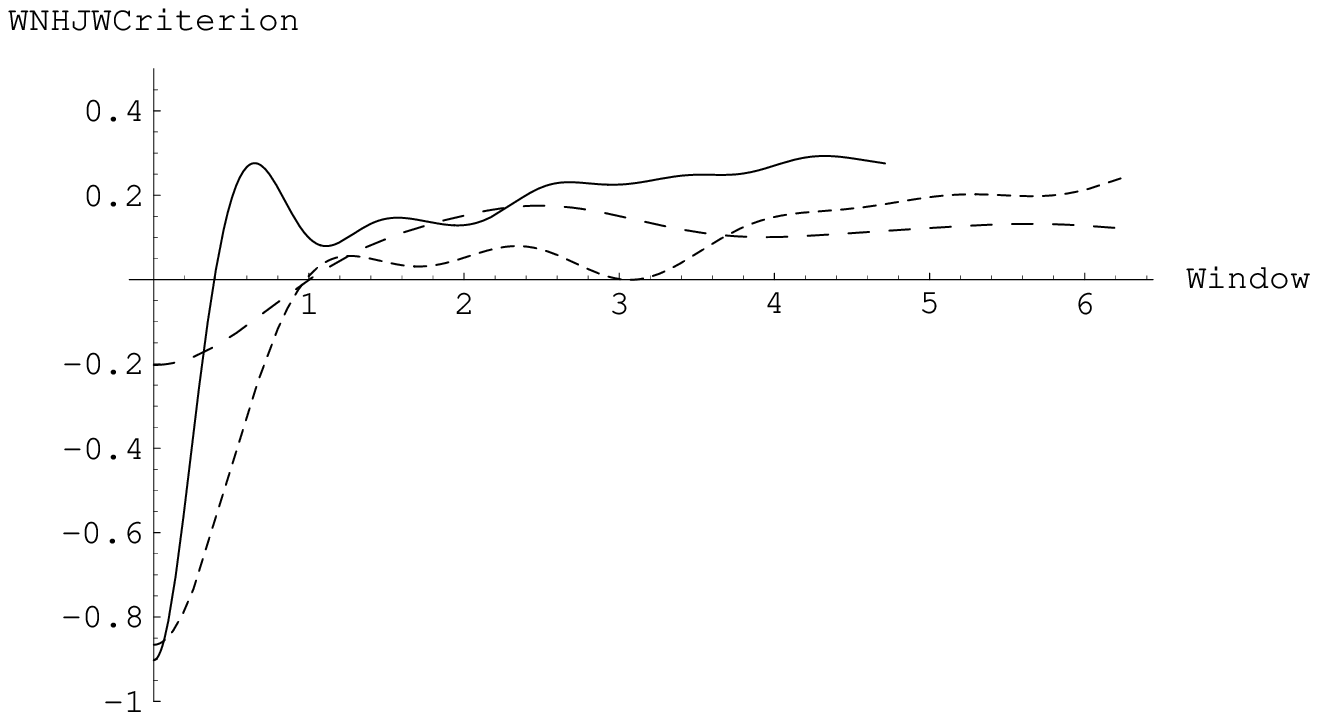} \caption {Time averages of
probabilities as functions of the time averaging window $2\Delta
t$. The time choices are those, at which the curves in the
previous figures are most negative, with the drawn curves
corresponding to those in the preceding figure. }
\label{DMLGIIntegr}
 \end{figure}

We solve \er{psiul}with the initial condition
$\psi_u(t=0)=1,\psi_l(0)=0$. Then the inequality in \er {LGW},
takes the simple form \beq |\psi_u(2t)|^2-|\psi_u(t)|^4\geq
0\label{LGW2}\enq The results for some choices of the parameters
in \er{H0} are shown in Figure \ref{DMLGAIInstProb} for an
infinitesimal averaging window, $2\Delta t\to 0$, corresponding to
an instantaneous or pure state. It is seen that for a small range
of time values [when the left hand side (LHS) of \er{LGW2} takes
negative values] the inequality is violated. Such violation may
either indicate the quantal nature of the system or to the
breakdown of the pre-conditions in \cite {WaldherrNHJW}. We next
 select some time values where each curve is most negative and
average the probabilities around these values, continuously
increasing the extent of the integration window. As seen in Figure
\ref{DMLGIIntegr}, above some finite size time windows the LGW
inequality becomes satisfied, though it was violated for an
infinitesimal time window (the pure state case). The transition
from negative to positive values may indicate the transition from
quantal to classical nature of the integrated DM in a continuous
manner, though this is also contingent to the satisfaction of the
"stability" hypothesis in \cite {WaldherrNHJW}. For the present
purposes the results demonstrate the facility with which
application of the time averaging procedure straddles the
quantal-classical gap.


\section{Conclusion}

 While the main result of this paper is the derivation of non-uniform weighting
 for the time coarsening process in section 6, a discussion of its application is appropriate.
  It may be argued that the Minimal Coupling Model
MCM used in this work is too specific to represent the wide
variety of real life interactions between a micro-system and its
environment. (Extensions of the model, possible tasks for future
work, are described in Appendix B.) However, it is almost
axiomatic that, excluding the small proportion of integrable
systems, in real life (exemplified in  \cite{LindenPSW} by the
Cooling Coffee Cup), thermal or equilibration processes depend
only in a minor way on the details of the micro-system-environment
coupling, and indeed most modelings of this coupling have hinged
on their amenability to solution, rather than on their being a
true description  of real life situations. The strength of the
time-averaging approach in this paper is the ease with which it
yields results (resolution of the time arrow problem being perhaps
the most obvious one); it also opens the way to explore concrete
physical situations by the selective, non-uniform time averaging,
as set out in section 5.

 \section {Acknowledgements} Thanks
are due to Rafael Ruppin for substantial help in the calculations
and to Ronnie Kosloff for a discussion.
\appendix
 \section{Original outline of  the three steps to
construct the density matrix }
 In the quoted previous work \cite
{EnglmanY2013} it was shown how the decohered-truncated phase of
the full subsystem-environment state (their DM), leading into the
reduced subsystem state, can be obtained by a time averaging
approach. The abstract steps needed to achieve this and the main
assumptions behind it were given by \cite{EnglmanY2013} in an
"Outline", which is briefly reproduced in this Appendix. The
Hamiltonians for which the system's states were determined were
(i) a Rabi-model (a single-spin in interaction with a single
vibrator), (ii) two mutually coupled spins, each coupled  to a
single vibrator. For completeness these Hamiltonians, formulated
in a semi-classical, time-dependent language, are also shown in
this article in Appendix B. While it is clear that a Hamiltonian
in which a time dependent term appears represents an {\it open}
system, what our results have shown is that even with the arguably
simplest form of the time dependent term (namely, a single
subsystem-environment harmonic interaction term), one obtains
results equivalent to those in the standard environment-tracing
formalism. The simplifying (and approximative) steps necessary to
pass from the Leggett {\it et al}'s (spin-oscillatory ensemble)
model \cite{LeggettEA} to ours were described in section 2.3 of
the earlier article. The present text has extended the
time-averaging formalism to issues not considered before,
including  second moments (fluctuations) of the DM (in section 6)
and its approach  through time averaging to classicality (section
7).
\begin{enumerate}

\item We adopt the von Neumann definition
%\cite{Neumann}-\cite{EnglmanY2004var}:
 \beq
 \rho_{ij}(t) = \frac{1}{(\sum _a 1)} \sum_{a} <i|\psi_{a}(t)><\psi_{a}(t)|j>
 \label{rho0}
 \enq In this definition the summation index $a$ represents the values of all coordinates, variables etc. external to the
 system (e.g., those of the environment affecting the system) and appearing also in the Hamiltonian.
 Thus the set ${\psi_{a}(t)}$ for all $a$'s forms a time dependent {\it ensemble} of states. The variables of the
 system themselves are implicit (not written out) in $\psi_a (t)$.

  \item We solve only for a single external condition  thus dispensing with the $a$ index in the wave function,
 but obtain $\rho(t)$ as the average over an adequate set of adjacent times:
 \beq
 \rho_{ij}(t)= \frac{1}{2 \Delta t}\int_{t-\Delta t}^{t+\Delta t}d\tau <i|\psi(\tau)>
 <\psi(\tau)|j>\label{rho2}
 \enq
  This should be equivalent to \er{rho0} if the ergodic hypothesis holds for the duration $2\Delta t$.
 While for several cases the off-diagonal matrix elements are small
 or vanishing, $\rho_{ij}(t)$, as defined above, represents in general a mixed state
 whose  diagonalized form
 $\rho_{ij}(t)\to \rho_{ii}^d(t)\delta_{ij}$  satisfies $\sum_i (\rho_{ii}^d(t))^2<1$ (section 5 in this paper).

  \end{enumerate}

\section {Minimal Coupling Hamiltonians}
(a) Single spin: \beq H(t)= e\sigma_z+k\sigma_x \sin(\omega t)
\label{HAO}\enq

(b) Interacting spins: \ber  \textbf{H}_{total}(t) & = & h(t) +
\textbf{H}_{int} \label {Htotal}\\h(t) & = &  \sum
_i[E_i\sigma_{zi} +k_i\sigma_{zi}\cos(\omega_i
t+\alpha'_i)+k'_i\sigma_{xi}\sin(\omega_i t+\alpha'_i)]
\label{h}\\\textbf{H}_{int} & = & \sum_{ij}[\gamma_{ij}
(\sigma_{zi} \cdot \sigma_{zj})+\gamma' (\sigma_{xi} \cdot
\sigma_{xj}+\sigma_{yi}\cdot \sigma_{yj})]\label{Hint}\enr having
written in the first line the total Hamiltonian, comprising the
parts in the next two lines. First, the spin energy terms in which
$E_i$ are energies of the spin systems, the $\sigma$'s are Pauli
matrices operating in the respective spin spaces; $k,\alpha$,
together with their tagged partners, are parameters of the
spin-boson couplings and $\omega$ is the frequency of the external
source. This external, boson source is classical and for it the
Hamiltonian need not be written out. The Hamiltonian does not
include back-reaction on the source, which can  at least
approximately be justified for periodic coupling and energies. The
last line is the spin-spin interaction term. A two-spin version of
this interacting-spin Hamiltonian was treated in \cite
{EnglmanY2013}.
 \section {Extremization of $G(t)$ in \er{G}}
We now prove that the maximum of $G(t)$ is unity when the wave
function components $\psi_u(t)$ and $\psi_l(t)$ are constant
during the $2\Delta t$ integration range \cite{trivial}. By
implication, $G(t)$ is less than unity (mixed state case) when the
components are genuinely time dependent. The method of proof is a
calculation of variations.

It has been stated in text that a maximum of $G(t)$ occurs when
the wave function components are throughout real and positive. We
further simplify by writing \beq \psi_u(t)\equiv
f(t),~~\psi_l(t)\equiv\sqrt{1-f^2(t)} \label {f}\enq Written as
\beq G[f(t)]=[\frac{1}{2\Delta t}\int_{t-\Delta t}^{t+\Delta
t}(2f^2(t')-1)dt']^2 +
 [\frac{1}{\Delta t}\int_{t-\Delta t}^{t + \Delta t}f(t')\sqrt(1-f^2(t'))dt']^2
\label{Gf} \enq a variation gives \ber \delta G[f(t)] & = &
[\frac{1}{\Delta t}\int_{t-\Delta t}^{t+\Delta
t}(2f^2(t')-1)dt'][\frac{1}{2\Delta t}\int_{t-\Delta t}^{t+\Delta
t}4f(t')\delta f(t')dt']\nonumber
\\ & + & [\frac{2}{\Delta t}\int_{t-\Delta t}^{t + \Delta
t}f(t')\sqrt{1-f^2(t')}dt'][\frac{1}{\Delta t}\int_{t-\Delta t}^{t
+ \Delta t}(\sqrt{1-f^2(t')}\nonumber\\
& - & \frac{f^2(t')}{\sqrt{1-f^2(t')}})\delta
f(t')dt']\label{DGf}\enr Remarkably, the cofactor of $\delta
f(t')$ vanishes when $f(t)=f$, a constant throughout the range of
integration, so that $G[f(t)=f]$ is then an extremum. Its  value
is unity, independent of $f$. That it is also a maximum can be
shown by evaluating $G(t)$ for the case that the function $f(t)$
takes two values $f_1,f_2\neq f_1$ in fractions $F, (1-F)$
(respectively) of the integration interval and showing that the
leading term in $G[f_1,f_2]$ is negative for $|f_1|,|f_2|\leq 1$,
namely \beq
-\frac{F(1-F)}{8}(f_1-f_2)^2[4+(f_1+f_2)^2][1+\frac{4+(f_1+f_2)^2}
{(4-(f_1+f_2)^2)^2}]\label{apprG}\enq
\section{Proof of Sum Formula in \er {F3} for a Finite
Bath.} It  is assumed that the subsystem and the large, but finite
sized bath form together a microcanonical ensemble with mean
energy of $E_{total}$ and small energy uncertainty $\nu$. The
range of the subsystem's energies $E(t)$, that vary in time, is
divided up into N segments, numbered $n(=1,...,N )$ each of spread
$\nu$. The segments contain  $r_{n+1}-r_n$ discrete (supposed
non-degenerate) energy levels $E^b_r$ of the bath, such that these
bath state energy levels satisfy \beq E_{total}= E(t)+
E^b_r\label{Etot}\enq for  some value of $E(t)[\approx E(t_n)]$
situated within the segment spread $\nu$. The corresponding time
spread over the segment is $t_{n+1}-t_n$, having assumed that for
the short passage time over the narrow segment spread the
subsystem energy is a monotonic, single-valued function of time.
This time spread is of the order of the bath's relaxation time.

The subsystem's density
 operator ("the reduced" density operator)
$\rho^S(t)$ is obtained in the bath tracing formalism by weighting
the system's proper density operator $\rho_n(t)$ in each segment
by the above number of bath states, giving \ber
\rho_n(t)(r_{n+1}-r_n) & = & \rho_n(t)\sum_{r_n}^{r_{n+1}}
\int_{E(t_n)}^{E(t_{n+1})}dE'\delta (E_{total}-E'-E^b_r)\nonumber\\
& = & \sum_{r_n}^{r_{n+1}} \rho_n(t)\int_{t_n}^{t_{n+1}}dt'
\frac{dE(t')}{dt'}\delta (E_{total}-E(t')-E^b_r)\nonumber\\
 & = & \sum_{r=r_n}^{r_{n+1}}\rho_n(t_r)\label{RDM}\enr having in the last line
 interchanged the order of the summation and the
time integration and taken heed of the property of the Dirac delta
function. We thus
 obtain for the weighted density operator in the time segment a discrete
sum with the time sum going over all such times that the system
energy is complemented to $E_{total}$ by a bath state's energy. In
the time-integrated formalism the system's density matrix is
obtained by integrating over time windows $t_w$ that are much
wider than the time spread of a segment. Then the resultant matrix
element is \beq \rho_{ij}\propto \sum_{t_r\epsilon t_w}
\rho_{ij}(t_r)\label{rhoij}\enq where the time sum goes over all
such times $t_r$ within the time window that the instantaneous
subsystem energy $E(t)$ is complemented by a bath level to add up
to $E_{total}$. The proportionality constant is fixed, by the
requirement that the trace of density matrix is unity. In the case
of an infinite bath with a continuous energy spectrum, the above
sum is replaced by a time integral including the energy {\it
density} of the bath levels $D^b(E^b)$: \beq \rho_{ij}(t)\propto
\int_{t}^{t+t_w}dt'D^b(E^b) \rho_{ij}(t')\label{rhoij2}\enq  This
formula is also applicable for the case that the bath is in
thermal equilibrium at a temperature $T$, so that the bath energy
density is proportional to \beq e^{\frac {E^b}{k_B T}}= e^{\frac
{E_{total}-E(t)}{k_B T}}\label{Gibbs}\enq by virtue of the
expression in \er{Etot} for the bath energy. This result,
originally due to \cite {RechtmanP}, has already been  used in
\cite {EnglmanY2013}, with the time independent factor $e^{\frac
{E_{total}}{k_B T}}$ having been absorbed in the proportionality
factor.

\end{document}